\magnification \magstep1
\raggedbottom
\openup 1\jot
\voffset6truemm
\def\cstok#1{\leavevmode\thinspace\hbox{\vrule\vtop{\vbox{\hrule\kern1pt
\hbox{\vphantom{\tt/}\thinspace{\tt#1}\thinspace}}
\kern1pt\hrule}\vrule}\thinspace}
\headline={\ifnum\pageno=1\hfill\else
\hfill {\it Axial Gauge in Quantum Supergravity} 
\hfill \fi}
\centerline {\bf AXIAL GAUGE IN QUANTUM SUPERGRAVITY}
\vskip 1cm
\centerline {Giampiero Esposito$^{1,2}$ and Alexander
Yu. Kamenshchik$^{3}$}
\vskip 1cm
\centerline {\it ${ }^{1}$Istituto Nazionale di Fisica
Nucleare, Sezione di Napoli,}
\centerline {\it Mostra d'Oltremare Padiglione 20, 
80125 Napoli, Italy;}
\centerline {\it ${ }^{2}$Dipartimento di Scienze Fisiche,}
\centerline {\it Mostra d'Oltremare Padiglione 19,
80125 Napoli, Italy;}
\centerline {\it ${ }^{3}$Nuclear Safety Institute, Russian
Academy of Sciences,}
\centerline {\it 52 Bolshaya Tulskaya, Moscow 113191, Russia.}
\vskip 1cm
\noindent
{\bf Abstract.} This paper studies the role of the axial gauge in
the semiclassical analysis of simple supergravity about
the Euclidean four-ball, when non-local boundary conditions
of the spectral type are imposed on gravitino perturbations
at the bounding three-sphere. Metric perturbations are instead
subject to boundary conditions completely invariant
under infinitesimal diffeomorphisms.
It is shown that the axial gauge
leads to a non-trivial cancellation of ghost-modes 
contributions to the one-loop divergence. The analysis, which
is based on zeta-function regularization, provides a full
$\zeta(0)$ value which coincides with the one obtained from
transverse-traceless perturbations for gravitons and gravitinos.
The resulting one-loop divergence does not vanish. This
property seems to imply that simple supergravity is not
even one-loop finite in the presence of boundaries.
\vskip 4cm
\noindent
To appear in Proceedings of the 12th Italian Conference on
General Relativity and Gravitational Physics
(Rome, September 1996).
\vskip 100cm
\leftline {\bf 1. Introduction}
\vskip 0.3cm
\noindent
When supergravity theories were introduced and the formalism
for their quantization was developed [1--3], there was the hope that
the supersymmetry relating bosonic and fermionic fields would
have improved the finiteness properties of pure gravity [4--6]
(although these theories are {\it not} perturbatively renormalizable).
However, the analysis of ultraviolet divergences is technically
so difficult that not even one-loop divergences were completely
analyzed in non-trivial backgrounds. More precisely, we are here
interested in the semiclassical analysis of quantum supergravity
in Riemannian four-manifolds with boundary. Boundary effects
are indeed crucial in the path-integral approach to quantum 
gravity and quantum cosmology [7,8], and the recent progress in the
quantization programme of field theories in the presence of 
boundaries has shed new light on the problems of Euclidean quantum
gravity [9--12] and on some properties of Euclidean Maxwell theory 
[13--16].

In the absence of boundaries, the massless gravitinos of simple
supergravity, with unrestricted gauge freedom, can only be studied 
in Ricci-flat backgrounds [17]. In the presence of boundaries,
however, one has to impose boundary conditions which involve both
the unprimed and the primed part of the gravitino potential
(in two-component spinor notation), hereafter denoted by
$\psi_{\mu}^{A}$ and ${\widetilde \psi}_{\mu}^{A'}$ respectively. 
Each of these spinor-valued one-forms admits a local description
in terms of a second potential provided that half of the conformal
curvature (i.e., self-dual or anti-self-dual) of the background 
vanishes. Thus, the Ricci-flat background is further restricted 
to be totally flat [18]. Our boundary three-geometry consists of one
or two three-spheres, since these are the spatial sections occurring
in the quantization of closed cosmological models.

As far as metric perturbations $h_{\mu \nu}$ are concerned, one
would like to impose boundary conditions written in terms of
projectors and/or first-order differential operators, completely
invariant under infinitesimal diffeomorphisms, and leading to a
self-adjoint second-order operator on $h_{\mu \nu}$. It has been
proved in Ref. [12] that this may be obtained by working in the axial
gauge (this being a {\it sufficient} condition to 
achieve self-adjointness) 
and hence setting all components of $h_{\mu \nu}$ equal to zero
at the boundary. 

For gravitino perturbations, one has a choice 
between non-local boundary conditions of the spectral type, and 
local boundary conditions motivated by local supersymmetry. In 
the former case, one sets to zero at the boundary half of
$\psi_{\mu}^{A}$ and ${\widetilde \psi}_{\mu}^{A'}$, i.e., 
those modes which multiply harmonics having positive eigenvalues
of the intrinsic three-dimensional Dirac operator of the 
boundary. This is a non-local operation, in that it relies on a
separation of the spectrum of such a three-dimensional elliptic
operator into its positive and negative parts. Local boundary
conditions for gravitinos involve instead complementary 
projectors, but they are {\it not} completely invariant under 
infinitesimal diffeomorphisms [19]. Hence we focus on the choice of 
spectral boundary conditions in the axial gauge, and we refer
the reader to Sec. IV of Ref. [20] for the analysis of local
boundary conditions.
\vskip 0.3cm
\leftline {\bf 2. Simple supergravity in the axial gauge:
one-loop results}
\vskip 0.3cm
\noindent
In the Faddeev-Popov path integral for the semiclassical
amplitudes of simple supergravity, one adds a gauge-averaging
term to the original Euclidean action for gravitons and
gravitinos, jointly with the corresponding 
ghost term [21]. In Ref.
[12] it has been shown that, on imposing the boundary conditions
$$
\Bigr[h_{ij}\Bigr]_{\partial M}
=\Bigr[h_{00}\Bigr]_{\partial M}
=\Bigr[h_{0i}\Bigr]_{\partial M}=0 
\; \; \; \; ,
\eqno (2.1)
$$
in the axial gauge, one obtains a unique, smooth and analytic
solution of the elliptic boundary-value problem, which picks
out transverse-traceless perturbations. Thus, when flat
Euclidean four-space is bounded by a three-sphere, the contribution
of gravitons to the one-loop divergence is [22]
$$
\zeta_{\rm TT}(0)=-{278\over 45} \; \; \; \; ,
\eqno (2.2)
$$
since gravitational ghost modes are forced to vanish everywhere
in the axial gauge [12].

For gravitinos, denoting by ${_{e}n_{CA'}}$ the two-spinor
version of the Euclidean normal to the boundary, the 
four-dimensional ghost operators in the axial gauge are found
to be [20]
$$
{\cal D}_{C}^{\; \; A} \equiv {_{e}n_{CC'}} \nabla^{AC'}
\; \; \; \; ,
\eqno (2.3)
$$
$$
{\cal F}_{C'}^{\; \; \; A'} \equiv {_{e}n_{CC'}} \nabla^{CA'}
\; \; \; \; .
\eqno (2.4)
$$
The corresponding ghost fields $\nu^{C}$ and $\mu^{C'}$ obey
Neumann conditions at the three-sphere boundary of radius $a$. 
These result from requiring
invariance of spectral boundary conditions for gravitino 
perturbations under infinitesimal gauge transformations.
Hence one finds the discrete spectra
$\lambda_{n}=(n+3)/a$, $\forall n \geq 0$, for
${\cal D}_{C}^{\; \; A}$, and ${\widetilde \lambda}_{n}=n/a$,
$\forall n \geq 0$, for ${\cal F}_{C'}^{\; \; \; A'}$. The
$\zeta(0)$ value for ghosts is thus found to be [20]
$$ \eqalignno{
\zeta(0)&=\zeta_{H}(-2,3)-3\zeta_{H}(-1,3)+2\zeta_{H}(0,3)
+2+\zeta_{R}(-2)+3\zeta_{R}(-1)+2\zeta_{R}(0) \cr
&=-{3\over 4}+{3\over 4}=0 \; \; \; \; ,
&(2.5)\cr}
$$
where $\zeta_{H}$ and $\zeta_{R}$ are the Hurwitz and 
Riemann zeta-functions, respectively.
Moreover, in the axial gauge, the four-dimensional elliptic
operator acting on Rarita-Schwinger potentials is
${\cal O}_{\; \; \nu}^{\mu} \equiv -\cstok{\ } 
\delta_{\; \; \nu}^{\mu}+{1\over \alpha}n^{\mu}n_{\nu}$.
Covariant differentiation of the resulting eigenvalue equation,
and contraction with flat-space $\gamma$-matrices, jointly with
spectral boundary conditions on $\psi_{\mu}^{A}$ and
${\widetilde \psi}_{\mu}^{A'}$, imply that a unique solution
exists, given by transverse-traceless perturbations. Their
contribution to $\zeta(0)$ is [23,24]
$$
\zeta_{{3\over 2}}(0)=-{289\over 360} \; \; \; \; ,
\eqno (2.6)
$$
and hence the full $\zeta(0)$ value turns out to be [20]
$$
\zeta(0)_{N=1 \; {\rm SUGRA}}=-{278\over 45}+{289\over 360}
=-{43\over 8} \; \; \; \; .
\eqno (2.7)
$$
It should be emphasized that the cancellation of contributions
of gauge and ghost modes in the axial gauge is a non-trivial
property in the presence of boundaries, since no such cancellation
occurs in covariant gauges. This has been proved by using both
geometric formulae for heat-kernel asymptotics (corresponding to
the Schwinger-DeWitt method) and explicit mode-by-mode calculations
[9--11,15]. Moreover, even in the Coulomb gauge for Euclidean
Maxwell theory such a cancellation was not found to occur in the
presence of boundaries [14].

When two concentric three-sphere boundaries of radii
$\tau_{-}$ and $\tau_{+}$ occur, which is the case more
relevant for quantum field theory, the equation obeyed by the
gravitino eigenvalues by virtue of spectral boundary conditions
turns out to be, for all $n \geq 0$ [20]
$$
I_{n+2}(M\tau_{+})K_{n+3}(M\tau_{-})
+I_{n+3}(M\tau_{-})K_{n+2}(M\tau_{+})=0
\; \; \; \; , 
\eqno (2.8)
$$
where $I_{n}$ and $K_{n}$ are the modified Bessel functions,
and $\tau_{+} > \tau_{-}$. This leads to [20]
$$
\zeta_{{3\over 2}}(0)=0 \; \; \; \; ,
\eqno (2.9)
$$
and hence the full one-loop divergence is given by [20]
$$
\zeta(0)_{N=1 \; {\rm SUGRA}}=-5 
\; \; \; \; ,
\eqno (2.10)
$$
since transverse-traceless graviton modes contribute $-5$,
and gravitino ghost modes vanish everywhere in this 
two-boundary problem [20].
\vskip 5cm
\leftline {\bf 3. Open problems}
\vskip 0.3cm
\noindent
The results (2.7) and (2.10) seem to imply that simple
supergravity is not even one-loop finite in the presence
of boundaries, when the axial gauge and spectral boundary
conditions are imposed. However, at least
three outstanding problems remain. They are as follows.
\vskip 0.3cm
\noindent
(i) One has to understand whether the reduction to
transverse-traceless perturbations for both gravitons and
gravitinos is a peculiar property of the axial gauge only,
within the framework of Faddeev-Popov formalism for quantum
amplitudes. Indeed, such a reduction has not been found to
occur in the case of local boundary conditions [20].
The relation between quantization schemes in different 
non-covariant gauges (e.g. axial vs. unitary) in the presence
of boundaries deserves careful consideration.
\vskip 0.3cm
\noindent
(ii) Higher-N supergravity models are naturally formulated in
backgrounds with a non-vanishing cosmological constant. How to
choose the boundary three-geometry ? How to deal with antisymmetric
tensor fields ? Is the one-loop divergence an unavoidable
feature of any problem with boundaries ?
\vskip 0.3cm
\noindent
(iii) How to study higher-order effects in perturbation theory
for simple or extended supergravity theories in the presence
of boundaries.

Maybe a new age is in sight in the understanding of ultraviolet
divergences in quantum supergravity. If this were the case, it
would shed new light on perturbative properties in quantum gravity 
and quantized gauge theories.
\vskip 0.3cm
\leftline {\bf Acknowledgements}
\vskip 0.3cm
\noindent
We are much indebted to Ivan Avramidi for scientific collaboration
on the role of the axial gauge in 
Euclidean quantum gravity (see Ref. [12]).
One of us (G.E.) is grateful to the INFN for financial support.
The work of A.Y.K. was partially supported by the Russian 
Foundation for Fundamental Researches through Grant No.
96-02-16220-a, and by the Russian Research Project
``Cosmomicrophysics".
\vskip 0.3cm
\leftline {\bf References}
\vskip 0.3cm
\item {[1]}
Freedman D. Z., van Nieuwenhuizen P. and Ferrara S. (1976)
{\it Phys. Rev. D} {\bf 13}, 3214.
\item {[2]}
Freedman D. Z. and van Nieuwenhuizen P. (1976) {\it Phys. Rev. D}
{\bf 14}, 912.
\item {[3]}
Kallosh R. E. (1978) {\it Nucl. Phys. B} {\bf 141}, 141.
\item {[4]}
't Hooft G. and Veltman M. (1974) {\it Ann. Inst. Henri Poincar\'e}
{\bf 20}, 69.
\item {[5]}
van Nieuwenhuizen P. and Vermaseren J. A. (1976) 
{\it Phys. Lett. B} {\bf 65}, 263.
\item {[6]}
van Nieuwenhuizen P. (1981) {\it Phys. Rep.} {\bf 68}, 189.
\item {[7]}
Hawking S. W. (1979) in {\it General Relativity, an Einstein
Centenary Survey}, edited by S. W. Hawking and W. Israel
(Cambridge University Press, Cambridge, 1979).
\item {[8]}
Hawking S. W. (1984) {\it Nucl. Phys. B} {\bf 239}, 257.
\item {[9]}
Esposito G., Kamenshchik A. Yu., Mishakov I. V. and
Pollifrone G. (1994) {\it Phys. Rev. D} {\bf 50}, 6329.
\item {[10]}
Esposito G., Kamenshchik A. Yu., Mishakov I. V. and
Pollifrone G. (1995) {\it Phys. Rev. D} {\bf 52}, 3457.
\item {[11]}
Esposito G. and Kamenshchik A. Yu. (1995) 
{\it Class. Quantum Grav.} {\bf 12}, 2715.
\item {[12]}
Avramidi I. G., Esposito G. and Kamenshchik A. Yu. (1996)
{\it Boundary Operators in Euclidean Quantum Gravity} 
(DSF preprint 96/10, HEP-TH 9603021, to appear in
{\it Class. Quantum Grav.}).
\item {[13]}
Esposito G. (1994) {\it Class. Quantum Grav.}
{\bf 11}, 905.
\item {[14]}
Esposito G. and Kamenshchik A. Yu. (1994) {\it Phys. Lett. B}
{\bf 336}, 324.
\item {[15]}
Esposito G., Kamenshchik A. Yu., Mishakov I. V. and Pollifrone G.
(1994) {\it Class. Quantum Grav.} {\bf 11}, 2939.
\item {[16]}
Esposito G., Kamenshchik A. Yu., Mishakov I. V. and Pollifrone G.
(1995) {\it Phys. Rev. D} {\bf 52}, 2183.
\item {[17]}
Deser S. and Zumino B. (1976) {\it Phys. Lett. B} {\bf 62}, 335.
\item {[18]}
Esposito G., Gionti G., Kamenshchik A. Yu., Mishakov I. V. 
and Pollifrone G. (1995) {\it Int. J. Mod. Phys. D}
{\bf 4}, 735.
\item {[19]}
Poletti S. (1990) {\it Phys. Lett. B} {\bf 249}, 249.
\item {[20]}
Esposito G. and Kamenshchik A. Yu. (1996) {\it One-Loop
Divergences in Simple Supergravity: Boundary Effects}
(DSF preprint 96/18, HEP-TH 9604182, to appear in
{\it Phys. Rev. D}).
\item {[21]}
Matsuki T. (1980) {\it Phys. Rev. D} {\bf 21}, 899.
\item {[22]}
Schleich K. (1985) {\it Phys. Rev. D} {\bf 32}, 1889.
\item {[23]}
D'Eath P. D. and Esposito G. (1991) {\it Phys. Rev. D}
{\bf 44}, 1713.
\item {[24]}
Kamenshchik A. Yu. and Mishakov I. V. (1992) 
{\it Int. J. Mod. Phys. A} {\bf 7}, 3713.

\bye